\documentclass[12pt]{article}
\usepackage{graphicx}
\title{Lorentz-invariant Bohmian description of inelastic scattering in QFT}
\author{Hrvoje Nikoli\'c \\
Theoretical Physics Division, Rudjer Bo\v{s}kovi\'{c} Institute, \\
P.O.B. 180, HR-10002 Zagreb, Croatia \\
{\normalsize hrvoje@thphys.irb.hr} \\
\makebox[1in]{} \\
}
\date{\today}
\begin{document}
\maketitle
\begin{abstract}
The Lorentz-invariant S-matrix elements in interacting
quantum field theory (QFT) are used to represent the QFT state
by a Lorentz-invariant many-time wave function. Such a wave function
can be used to describe inelastic scattering processes
(involving particle creation and destruction) by Bohmian particle trajectories 
satisfying relativistic-covariant equations of motion. 
\end{abstract}

\vspace*{0.5cm}

\section{Introduction}

The Bohmian formulation of quantum mechanics (QM) 
\cite{BMbook1,BMbook2,BMbook3,BMbook4,BMbook5,BMbook6}
allows to visualize quantum processes in terms of continuous deterministic
pointlike-particle trajectories guided by wave functions. 
In this formulation, all quantum uncertainties
emerge from ignorance of the actual initial particle positions. 
Such a formulation not only offers a possible answer to deep conceptual and interpretational 
puzzles of QM, but in some cases provides also a powerful practically useful computational
tool \cite{BMbook3,lopr}.

A challenge for the Bohmian formulation of quantum theory is to reconcile 
continuous particle trajectories with phenomena of particle creation and destruction
in quantum field theory (QFT).
One possibility is to explicitly break the rule of continuous deterministic evolution,
by adding an additional equation that specifies stochastic breaking of the trajectories
\cite{durrcr1,durrcr2}. Another possibility is to introduce an additional
continuously and deterministically evolving hidden variable that specifies
effectivity of each particle trajectory \cite{nikcr1,nikcr2}. However,
both possibilities seem rather artificial and contrived.
In addition, the explicit constructions in \cite{durrcr1,durrcr2,nikcr1,nikcr2}
do not obey Lorentz invariance. 

Recently, a much simpler approach
has been introduced \cite{nikolQFT}, in which the particle trajectories are
continuous, but the appearance of particle creation and destruction 
results from quantum theory of measurements
describing entanglement with particle detectors.
The QFT states are represented by many-time wave functions
(which, as a byproduct, offers a new resolution
of the black-hole information paradox \cite{nikbh}),
implying that the corresponding Bohmian equations of motion
for particle trajectories are relativistic-covariant. The equations
determining the many-time wave functions are explicitly Lorentz-covariant
for free fields. In the interacting case, however, the many-time evolution
of wave functions in \cite{nikolQFT} is described by the time-evolution operators
$\hat{U}(t)$ in the interaction picture, which lack manifest Lorentz invariance
for interacting QFT.

In this paper we further develop some of the ideas introduced in \cite{nikolQFT}.
In particular, we find a new method for calculation of the many-time wave function
in interacting QFT, which turns out to be (i) simpler than that in
\cite{nikolQFT} and (ii) manifestly Lorentz-invariant.
In addition, we propose a new guiding equation for Bohmian trajectories
which turns out to be much simpler than that in \cite{nikolQFT}.
We also clarify some conceptual issues in a somewhat different
(and hopefully more illuminating) way than in \cite{nikolQFT},
and discuss some limitations of the present approach that represent a challenge
for the future research.

The paper is organized as follows. In Sec.~\ref{SEC2} we present a brief review
of the Bohmian formulation of relativistic QM, as well as a general qualitative description
of particle creation and destruction, with emphasis on conceptual (rather than technical)
aspects. In Sec.~\ref{SEC3} we describe in detail how states in free and interacting
QFT can be represented by Lorentz-invariant many-time wave functions.
In Sec.~\ref{SEC4} we give a probabilistic interpretation of these wave functions
and develop the corresponding Bohmian interpretation in terms of particle trajectories.
The conclusions are drawn in Sec.~\ref{SEC5}.

In the paper we use units $\hbar=c=1$ and the metric signature $(+,-,-,-)$.

\section{Conceptual preliminaries}
\label{SEC2}

\subsection{Outline of relativistic Bohmian mechanics}
\label{SEC2.1}

Since Bohmian mechanics (BM) is a nonlocal theory, it is frequently
objected that it is not compatible with relativity.
Yet, as reviewed in \cite{niktorino}, all objections
of that type can be circumvented.
Relativistic-covariant nonlocal Bohmian equations of motion 
for particle trajectories have been
introduced in \cite{durr96} and further studied in
\cite{nik3,nik4,hern}. Using the idea that, in relativistic QM,
space probability density should be generalized to spacetime
probability density (see, e.g., \cite{broyles,niktime}), 
a relativistic-invariant probabilistic interpretation
(associated with relativistic particle trajectories)
has been introduced in \cite{niktime,nikbosfer}.
As indicated in \cite{nikolQFT} and generally shown in \cite{niktorino},
this makes the measurable statistical predictions of relativistic BM
compatible with {\em all} measurable statistical predictions of the
``ordinary" purely probabilistic interpretation of QM.

Let us present a brief overview of the main ideas of relativistic BM.
Denoting a point in spacetime by $x=\{ x^{\mu} \}$, 
an $n$-particle quantum state is represented by a many-time wave function
$\psi(x_1,\ldots,x_n)$. 
This wave function lives in the $4n$-dimensional configuration space
with coordinates $x^{\mu}_a$, $a=1,\ldots,n$, referred to as
{\em relativistic configuration space}.
This is a many-time wave function \cite{tomon},
because each particle has its own time-coordinate $x^0_a$.
The quantity $\psi^*\psi$ is interpreted as probability density 
in the relativistic configuration space, in the sense that
the infinitesimal probability $dP$ of finding $n$ particles in an 
infinitesimal $4n$-dimensional volume around the points $x_1,\ldots,x_n$
is given by
\begin{equation}\label{RBM1}
dP = \psi^*(x_1,\ldots,x_n)\psi(x_1,\ldots,x_n) 
d^4x_1 \cdots d^4x_n .
\end{equation}
The wave function satisfies the $n$-particle Klein-Gordon equation
\begin{equation}\label{KGn}
\sum_{a=1}^n [\partial_a^{\mu}\partial_{a\mu}+m_a^2]\psi=0 , 
\end{equation}
which implies that the $n$-particle Klein-Gordon current
\begin{equation}\label{J_a}
J^{\mu}_a=  
\frac{i}{2} \psi^*\!\stackrel{\leftrightarrow\;}{\partial^{\mu}_a}\! \psi  
\end{equation}
is conserved:
\begin{equation}\label{Jcons}
\sum_{a=1}^{n}\partial_{a\mu}J^{\mu}_a =0 .
\end{equation}

The Bohmian particle trajectories are integral
curves of the vector field $V^{\mu}_a(x_1,\ldots,x_n)$ calculated from the wave function as
\begin{equation}\label{RBM2}
V^{\mu}_a= \frac{J^{\mu}_a}{\psi^*\psi}. 
\end{equation}
In the parameterized form, the integral curves are represented by functions
$X^{\mu}_a(s)$ satisfying
\begin{equation}\label{RBM3}
\frac{dX^{\mu}_a(s)}{ds} = V^{\mu}_a(X_1(s),\ldots,X_n(s)) .
\end{equation}
From (\ref{Jcons}) and the fact that $\psi$ 
and does not have an explicit dependence on $s$, one finds the
equivariance equation
\begin{equation}\label{RBM4}
\frac{\partial (\psi^*\psi)}{\partial s} + 
\sum_{a=1}^{n}\partial_{a\mu}(\psi^*\psi V^{\mu}_a ) =0 .
\end{equation}
Formally, Eq. (\ref{RBM4}) shows that (\ref{RBM3}) is compatible with
(\ref{RBM1}) in the following sense: If a statistical ensemble
of particles has the probabilistic distribution
(\ref{RBM1}) for some initial $s$, then it has the 
probabilistic distribution (\ref{RBM1}) for {\em any} $s$.

To understand the physical meaning of the formal equivariance equation (\ref{RBM4}),
one needs to understand the physical meaning of the parameter $s$.
For that purpose, it is useful to exploit
the analogy with nonrelativistic Newtonian mechanics \cite{niksuperlum}.
For nonrelativistic particle systems with conserved energy, 
the forces do not
have an explicit dependence on time $t$. The {\em only} quantities
that have a dependence on $t$ are particle trajectories
$X^i_a(t)$, $i=1,2,3$. Thus, the parameter $t$ has a physical meaning only
along trajectories; time without trajectories does not exist! 
In this sense, $t$ is only an auxiliary parameter that serves
to parameterize the trajectories in 3-dimensional space, 
not a fundamental physical quantity by its own. Yet, a ``clock" can measure time {\em indirectly}. Namely, a ``clock" is nothing
but a physical process described by a function $X^i_a(t)$ periodic
in $t$. One actually observes the number of periods, 
and then interprets it as a measure of elapsed time.

The theory of relativity revolutionized the concept of time
by replacing the parameter $t$ with a coordinate $x^0$ treated
as a 4th dimension not much different from other 3 space dimensions.
Yet, it does not mean that an auxiliary Newton-like time parameter
is completely eliminated from relativistic mechanics.
Such a parameter can still be introduced to parameterize
relativistic spacetime particle trajectories in a manifestly 
covariant manner. 
This parameter, denoted by $s$, can be identified
with a generalized proper time defined along particle trajectories
of many-particle systems \cite{niksuperlum}. 
The parameter $s$ can even be measured indirectly by a ``clock" corresponding 
to a physical process periodic in $s$, 
in complete analogy with measurement of $t$ in nonrelativistic mechanics.
As discussed in more detail in \cite{niksuperlum},
this makes the parameter $s$ appearing in (\ref{RBM3}) 
a physical quantity, very much analogous to Newton time $t$.
With this physical insight, the relativistic-covariant equation
(\ref{RBM4}) is to be interpreted
as physical probability conservation during the evolution
parameterized by the evolution-parameter $s$.

Now let us sketch how all statistical predictions of the 
purely probabilistic interpretation of QM can be reproduced
from relativistic BM (for more details see \cite{niktorino}). 
Let a physical system be described
by a wave function
\begin{equation}\label{nikolic:meas0}
 \psi(x)=\sum_b c_b \psi_b(x) ,
\end{equation}
where $\psi_b(x)$ are eigenstates of some hermitian operator $\hat{B}$ 
on the Hilbert space
of functions of $x$, normalized 
in a large but finite 4-dimensional box such that 
$\int d^4x \, \psi^*_b(x)\psi_b(x) =1$.
The purely probabilistic interpretation asserts that
$|c_b|^2$ is the probability that the observable 
$B$ will take the value $b$. To see how BM reproduces this
assertion, one needs to take into account the entanglement
with the measuring apparatus. This leads to a wave function \cite{niktorino}
\begin{equation}\label{nikolic:meas2}
 \psi(x,y) = \sum_b c_b \psi_b(x)E_b(y) ,
\end{equation}
where $E_b(y)$ are detector wave functions normalized in
the relativistic configuration space, with a negligible 
overlap in that space. Since (\ref{RBM3}) is compatible with (\ref{RBM1}),
the probability that detector particles will have a relativistic 
configuration $Y$ in the support of $E_b(y)$ is equal to $|c_b|^2$ 
\cite{niktorino}. This shows that the measurable statistical predictions
of relativistic BM coincide with those of purely probabilistic 
interpretation. 

In particular, if $\psi_b(x)$ are (approximate) eigenstates of the space-position operators
$x^1$, $x^2$, $x^3$ at time $x^0$, then the measurement procedure above 
(approximately) reproduces
the usual space probability distribution $dP_{(3)}=\psi^*(x)\psi(x) d^3x$. 

\subsection{The general mechanism of particle creation and destruction}
\label{SEC2.2}

In Sec.~\ref{SEC2.1} we have studied relativistic QM with a fixed number
$n$ of particles. Now we give a qualitative description
of the general mechanism of particle creation and destruction
in QFT. Schematically, it can be described in 3 steps.

In the {\it first step}, the initial state $|n_{\rm initial}\rangle$
with a definite initial number of particles $n_{\rm initial}$ suffers a
unitary deterministic evolution in interacting QFT
\begin{equation}\label{step1}
|n_{\rm initial}\rangle \rightarrow \displaystyle\sum_{n} c_{n}|n\rangle ,
\end{equation}
where the final state is a superposition of states $|n\rangle$ with different
numbers of particles.

In the {\it second step}, the quantum state above interacts with the environment
(e.g., a particle detector), which causes a unitary evolution that creates
entanglement with environment states $|E_{n}\rangle$: 
\begin{equation}\label{step2}
 \left[ \sum_{n} c_{n}|n\rangle \right]  |E_{\rm initial} \rangle \rightarrow 
\sum_{n} c_{n}|n\rangle |E_{n}\rangle .
\end{equation}
Here state $|E_{n}\rangle$ can be thought of as a macroscopic state 
describing a detector in a state of saying that $n$ particles are detected.
Since different macroscopic states are macroscopically distinguishable, the corresponding
wave functions $E_{n}(y)$ have a negligible overlap:
\begin{equation}\label{overlap}
 E_{n}(y)E_{n'}(y) \simeq 0 \;\;\; {\rm for} \;\;\; n\neq n' .
\end{equation}

In the {\it third step}, one needs a mechanism that will pick up only one term
in the superposition (\ref{step2}). Conventionally, it is usually described by the 
wave-function ``collapse''. The role of the Bohmian formulation
is to replace this {\it ad hoc} collapse with a mathematically better 
defined physical process. The wave function depending on $y$
guides the detector particles with a trajectory $Y(s)$. 
Due to (\ref{overlap}), 
the particles enter only one channel $E_{n_{\rm final}}(y)$
among many channels $E_{n}(y)$ in  (\ref{step2}).
This makes all other channels empty, which for all practical purposes is
effectively the same as if the state exhibited a collapse
\begin{equation}\label{step3}
 \sum_{n} c_{n}|n\rangle |E_{n}\rangle \rightarrow |n_{\rm final}\rangle |E_{n_{\rm final}}\rangle .
\end{equation}
(For more details see also \cite{nikolQFT}.)

The effect of these 3 steps can be summarized as a transition
\begin{equation}
|n_{\rm initial}\rangle |E_{\rm initial} \rangle \rightarrow 
|n_{\rm final}\rangle |E_{n_{\rm final}}\rangle ,
\end{equation}
which typically involves the destruction of some initial particles and the creation of
some new (final) ones.

\section{QFT states represented by Lorentz-invariant wave functions}
\label{SEC3}

\subsection{Momentum-eigenstates with fixed number of particles}
\label{SECfixedn}

Free QFT is usually formulated in terms of
$n$-particle states $|k_1,\ldots,k_n\rangle$
with on-shell 4-momenta $k_a$, $a=1,\ldots, n$.
(For simplicity, we consider particles without spin, but spin can also be included
as in \cite{nikolQFT}.)
Introducing the condensed notation $\vec{k}^{(n)} = (k_1,\ldots,k_n)$, 
we denote these states as
\begin{equation}
 |n,\vec{k}^{(n)}\rangle \equiv |k_1,\ldots,k_n\rangle . 
\end{equation}
The state $|n,\vec{k}^{(n)}\rangle$ can also be
represented by an $n$-point wave function \cite{nikolQFT}
\begin{eqnarray}\label{wfnk}
\psi_{n,\vec{k}^{(n)}}(x_1,\ldots,x_n) & = & \langle x_1,\ldots,x_n |k_1,\ldots,k_n \rangle
\nonumber \\
& = & b_n S_{\{x_1,\ldots,x_n\}}  e^{-ik_1 x_1}\cdots e^{-ik_n x_n},
\end{eqnarray}
where $b_n$ is a normalization factor and $ S_{\{x_1,\ldots,x_n\}}$ denotes symmetrization
over all $x_1,\ldots,x_n$. 
The normalization factor $b_n$  is chosen such that \cite{nikolQFT}
\begin{equation}
 \int {\cal D}\vec{x}^{(n)} |\psi_{n,\vec{k}^{(n)}}(\vec{x}^{(n)})|^2 =1 ,
\end{equation}
where $\vec{x}^{(n)} = (x_1,\ldots,x_n)$ and
\begin{equation}
 {\cal D}\vec{x}^{(n)} = d^4x_1 \cdots d^4x_n .
\end{equation}

\subsection{General states in free QFT}
\label{SECuncertn}

A general free QFT state has an expansion of the form
\begin{equation}\label{Psi}
 |\Psi\rangle = c_0|0\rangle +
\sum_{n=1}^{\infty} \sum_{\vec{k}^{(n)}} c_{n,\vec{k}^{(n)}} |n,\vec{k}^{(n)}\rangle ,
\end{equation}
where $|0\rangle$ is the vacuum. We assume that the state is normalized to unity, 
in the sense that
\begin{equation}
\langle \Psi |\Psi\rangle = |c_0|^2 + 
\sum_{n=1}^{\infty} \sum_{\vec{k}^{(n)}} |c_{n,\vec{k}^{(n)}}|^2 =1 .
\end{equation}

Since (\ref{Psi}) involves a superposition
of states with {\em various} numbers $n$ of particles, it is convenient to slightly
modify the notation of Sec.~\ref{SECfixedn}.
To distinguish particle positions $x_a$ coming from sectors of different $n$,
instead of $\vec{x}^{(n)} = (x_1,\ldots,x_n)$ we write 
\begin{equation}
 \vec{x}^{(n)} \equiv (x_{n,1},\ldots,x_{n,n}) .
\end{equation}
Then the collection of particle positions from sectors of {\em all} possible $n$ is denoted as
\begin{equation}\label{vecx}
 \vec{x}=(\vec{x}^{(1)},\vec{x}^{(2)},\ldots ) =
(x_{1,1}, x_{2,1},x_{2,2},\ldots ) .
\end{equation}
The state (\ref{Psi}) can be represented by a many-component wave function
\begin{equation}\label{Psix}
\Psi(\vec{x}) = 
\left( 
\begin{array}{c} 
 \tilde{\Psi}_0 \\ \tilde{\Psi}_1(\vec{x}^{(1)}) 
\\ \tilde{\Psi}_2(\vec{x}^{(2)}) \\ \vdots
\end{array}
\right) .
\end{equation}
Here
\begin{equation}\label{e0}
\tilde{\Psi}_0 = \sqrt{ \frac{1}{{\cal V}} } \, c_0 ,
\end{equation}
\begin{equation}\label{en}
\tilde{\Psi}_n(\vec{x}^{(n)}) = \sqrt{ \frac{{\cal V}^{(n)}}{{\cal V}} } 
\tilde{\psi}_n(\vec{x}^{(n)}) ,
\end{equation}
where 
\begin{equation}
 {\cal V}^{(n)}=\int {\cal D}\vec{x}^{(n)} ,
\;\;\;\;\; {\cal V}=\prod_{n=1}^{\infty}{\cal V}^{(n)} , 
\end{equation}
are volumes of the corresponding relativistic configuration spaces, and
\begin{equation}
\tilde{\psi}_n(\vec{x}^{(n)}) = \sum_{\vec{k}^{(n)}} c_{n,\vec{k}^{(n)}} 
\psi_{n,\vec{k}^{(n)}} (\vec{x}^{(n)}) 
\end{equation}
are $n$-particle wave packets with $\psi_{n,\vec{k}^{(n)}} (\vec{x}^{(n)})$ given by
(\ref{wfnk}).
The tilde above wave functions denotes wave functions which are not necessarily normalized
to unity. The normalization factors in (\ref{e0})-(\ref{en}) are chosen such that
\cite{nikolQFT}
\begin{equation}
\int {\cal D}\vec{x} \, |\tilde{\Psi}_0|^2 = |c_0|^2 , \;\;\;\;\;
 \int {\cal D}\vec{x} \, |\tilde{\Psi}_n|^2 = \sum_{\vec{k}^{(n)}} |c_{n,\vec{k}^{(n)}}|^2 ,
\end{equation}
where
\begin{equation}
 {\cal D}\vec{x} = \prod_{n=1}^{\infty} {\cal D}\vec{x}^{(n)} .
\end{equation}
This provides that the total wave function (\ref{Psix}) is normalized to unity, in the sense that
\begin{equation}
\int {\cal D}\vec{x} \, \Psi^{\dagger}(\vec{x})\Psi(\vec{x}) =
\int {\cal D}\vec{x} \, |\tilde{\Psi}_0|^2 + 
\sum_{n=1}^{\infty} \int {\cal D}\vec{x} \, |\tilde{\Psi}_n|^2 = 1 .
\end{equation}

Thus we see that the many-time wave function in free QFT is uniquely determined by a set of
expansion coefficients $c_0$, $c_{n,\vec{k}^{(n)}}$.

\subsection{Scattering wave function}

In interacting QFT, particles may be created and destructed. The information on dynamics
of this creation and destruction in encoded in the scattering matrix (shortly, S-matrix)
with matrix elements in the momentum space (see, e.g., \cite{BD2,ryder,kaku})
\begin{equation}\label{Smatrix}
\langle k_1,\ldots,k_n|\hat{S}|p_1,\ldots,p_m \rangle \equiv
S(n,\vec{k}^{(n)}| m,\vec{p}^{(m)}) .
\end{equation}
These matrix elements are Lorentz-invariant. 
In general, the initial state before scattering is a superposition of the form (\ref{Psi})
with expansion coefficients
$c_0^{\rm in}$, $c_{n,\vec{k}^{(n)}}^{\rm in}$.
Then, the final state after  scattering is also a superposition of the form (\ref{Psi}),
but with different expansion coefficients
$c_0^{\rm out}$, $c_{n,\vec{k}^{(n)}}^{\rm out}$
given by
\begin{equation}\label{transition}
 c_{n,\vec{k}^{(n)}}^{\rm out} = \sum_m \sum_{\vec{p}^{(m)}}
S(n,\vec{k}^{(n)}| m,\vec{p}^{(m)})  \; c_{m,\vec{p}^{(m)}}^{\rm in} .
\end{equation}
(Eq. (\ref{transition}) can be thought of as including the vacuum terms with
$n=0$ and $m=0$, but they are not written explicitly because
the S-matrix elements involving vacuum are usually trivial.)
Thus, the total wave function can be written as
\begin{equation}\label{scatPsi}
\Psi(\vec{x}) \simeq 
\left\{ 
\begin{array}{l} 
\displaystyle\sqrt{ \frac{{\cal V}}{{\cal V}^{\rm in}} }
\Psi^{\rm in}(\vec{x}) \;\;\;\; {\rm for} \; \vec{x} \in {\cal R}^{\rm in}, \\
\displaystyle\sqrt{ \frac{{\cal V}}{{\cal V}^{\rm out}} } 
\Psi^{\rm out}(\vec{x}) \;\;\;\; {\rm for} \; \vec{x} \in {\cal R}^{\rm out} ,
\end{array}
\right. 
\end{equation} 
where ${\cal R}^{\rm in}$ and ${\cal R}^{\rm out}$ are in and out regions 
of the relativistic configuration space with volumes 
${\cal V}^{\rm in}$ and ${\cal V}^{\rm out}$, respectively.
The wave functions $\Psi^{\rm in}(\vec{x})$ and $\Psi^{\rm out}(\vec{x})$
are defined as in Sec.~\ref{SECuncertn}, with coefficients
$c_0^{\rm in}$, $c_{n,\vec{k}^{(n)}}^{\rm in}$ and
$c_0^{\rm out}$, $c_{n,\vec{k}^{(n)}}^{\rm out}$, respectively.
The normalization factors in (\ref{scatPsi}) are chosen such that
$\int {\cal D}\vec{x} \, \Psi^{\dagger}(\vec{x})\Psi(\vec{x}) = 1$. 

Let us make a few comments on validity of the approximation (\ref{scatPsi}).
Strictly speaking, the S-matrix elements (\ref{Smatrix}) refer only to
in states $|p_1,\ldots,p_m \rangle$ at $x^0_a\rightarrow -\infty$ and
out states  $|k_1,\ldots,k_m \rangle$ at $x^0_a\rightarrow \infty$.
Nevertheless, they also represent a good approximation for large
but finite in and out times. After all, the predictions obtained
from these matrix elements are in excellent agreement with experiments,
and no experiments are performed at infinity. In fact, the S-matrix 
elements represent a good approximation wherever particles can be 
well approximated by free particles. But this means that 
the approximation (\ref{scatPsi}) is good {\em almost everywhere},
except in a very small region of spacetime at which the collision
of localized wave packets actually happens. For relativistic collisions, 
the size of this small region of collision is typically
of the order $1/E$, where $E$ is a typical energy of the 
colliding particles. (For decay processes, 
$E$ is the decay width, which is large for short-living particles.)
Thus, we can conclude that, in a typical situation,
(\ref{scatPsi}) is a good approximation almost
everywhere, i.e., that ${\cal R}^{\rm in} \cup {\cal R}^{\rm out}$
covers almost the whole relativistic configuration space.

Let us compare it with the results of \cite{nikolQFT}. In principle, the method developed
in \cite{nikolQFT} allows to calculate $\Psi(\vec{x})$ everywhere,
but is not manifestly Lorentz-invariant. By contrast, the method developed here
is manifestly Lorentz-invariant, but is not valid everywhere. It seems that
it should be possible to develop a method that is both 
valid everywhere and manifestly Lorentz-invariant, but we leave 
the development of such a method as a program for the future research.

\section{Probability and particle trajectories in QFT}
\label{SEC4}

In this section we develop the physical interpretation of the wave function
$\Psi(\vec{x})$, as a natural generalization of the results in Sec.~\ref{SEC2.1}.
The probabilistic interpretation (\ref{RBM1}) generalizes to
\begin{equation}\label{Prob}
 {\cal D}P=\Psi^{\dagger}(\vec{x})\Psi(\vec{x}) \, {\cal D}\vec{x}.
\end{equation}
Due to (\ref{Psix}), the probability density $\Psi^{\dagger}\Psi$ decomposes as
\begin{equation}\label{decomp}
\Psi^{\dagger}(\vec{x})\Psi(\vec{x}) =|\tilde{\Psi}_0|^2 +
\sum_{n=1}^{\infty} |\tilde{\Psi}_n(\vec{x}^{(n)})|^2 .
\end{equation}
Each $n$-particle wave function $\tilde{\Psi}_n(\vec{x}^{(n)})$ satisfies the $n$-particle Klein-Gordon equation of the form of (\ref{KGn}).

For further analysis, it is convenient to introduce a condensed label $A=(n,a_n)$, such that 
(\ref{vecx}) can be written as
\begin{equation}
\vec{x}=\{ x_A \} = (x_1,x_2,x_3,\ldots) . 
\end{equation}
With this notation, we introduce the current
\begin{equation}\label{J_A}
J^{\mu}_A = \frac{i}{2} \Psi^{\dagger}\!\stackrel{\leftrightarrow\;}{\partial^{\mu}_A}\! \Psi ,
\end{equation}
which generalizes (\ref{J_a}).
Due to (\ref{decomp}), it can be decomposed into a collection of $n$-particle currents,
each of which is conserved due to the $n$-particle Klein-Gordon equation.
This implies that (\ref{J_A}) is also conserved 
\begin{equation}
 \sum_{A=1}^{\infty} \partial_{A\mu} J^{\mu}_A =0 , 
\end{equation}
which implies a generalization of (\ref{RBM4})
\begin{equation}\label{RBM4gen}
\frac{\partial (\Psi^{\dagger}\Psi)}{\partial s} + 
\sum_{A=1}^{\infty}\partial_{A\mu}(\Psi^{\dagger}\Psi V^{\mu}_A ) =0 ,
\end{equation}
where 
\begin{equation}
V^{\mu}_A=\frac{J^{\mu}_A}{\Psi^{\dagger}\Psi} 
\end{equation}
generalizes (\ref{RBM2}).
The Bohmian particle trajectories are given by the generalization of (\ref{RBM3}) 
\begin{equation}\label{RBM3'}
\frac{dX^{\mu}_A(s)}{ds} = V^{\mu}_A(X_1(s),X_2(s),X_3(s),\ldots) ,
\end{equation} 
which are compatible with (\ref{Prob}) due to the equivariance equation
(\ref{RBM4gen}).

It is straightforward to apply this to the scattering wave function (\ref{scatPsi}),
which, together with physical insights from Sec.~\ref{SEC2.2},
provides a Bohmian description of inelastic scattering processes involving particle
creation and destruction.

\section{Conclusion}
\label{SEC5}

The Bohmian formulation of quantum theory describes all quantum processes
in terms of continuous deterministic particle trajectories guided by wave functions.
The results of the present paper show that it can be 
formulated in a form which (i) obeys manifest Lorentz invariance and (ii) includes
a description of particle creation and destruction in QFT. (For simplicity, in this paper
we have discussed only particles without spin, but, by using the results
presented in \cite{nikolQFT}, the generalization to particles with spin  
is relatively simple.) These results reinforce the view that the Bohmian formulation is a 
viable formulation of quantum theory.

\section*{Acknowledgements}

This work was supported by the Ministry of Science of the
Republic of Croatia under Contract No.~098-0982930-2864.

\end{document}